\documentclass[conference,a4paper]{IEEEtran}
\IEEEoverridecommandlockouts 
\usepackage{cite}
\usepackage{svg}
\usepackage{amsmath,amssymb,amsfonts}
\usepackage[ruled,vlined,linesnumbered]{algorithm2e} 
\usepackage{algpseudocode}
\usepackage{tabularx}
\usepackage{dblfloatfix} 
\usepackage{graphicx}
\usepackage{textcomp} 
\usepackage{xcolor}
\usepackage{floatrow}
\usepackage[skip=0pt]{caption}
\usepackage{etoolbox}
\DeclareGraphicsExtensions{.pdf,.jpeg,.png,.jpg,.bmp,.tif}  
\usepackage{subcaption} 
\usepackage{multicol}  
\begin{document} 
\title{Adaptive Digital Twin and Communication-Efficient Federated Learning Network Slicing for 5G-enabled Internet of Things}
\makeatletter
\makeatother
 \author{
\IEEEauthorblockN{Daniel Ayepah-Mensah}
\IEEEauthorblockA{School of Computer Science and Engineering,\\
University of Electronic Science \\ and Technology of China, UESTC\\
Chengdu, China\\
Email:ayeps2000@gmail.com}
  \and
\IEEEauthorblockN{ Guolin~Sun}
\IEEEauthorblockA{School of Computer Science and Engineering,\\
University of Electronic Science \\ and Technology of China, UESTC\\
Chengdu, China\\
Email: gulin.sun@uestc.edu.cn}
  \and
  \IEEEauthorblockN{Yu Pang}
  \IEEEauthorblockA{\hspace{15mm} School of Electro-optical Engineering, \\
\hspace{10mm} Chongqing  University of Posts and Telecommunications,  CUPT,\\
 \hspace{10mm} Chongqing, China\\
\hspace{15mm} Email: pangyu@cqupt.edu.cn}
  \and
  \IEEEauthorblockN{Wei Jiang}
  \IEEEauthorblockA{Department of Electrical  and Information  \\ Technology, \\ 
Technische University, TU,\\
Kaiserslautern, Germany\\
Email: wei.jiang@dfki.de}}

\maketitle 
\begin{abstract}
Network slicing enables industrial Internet of Things (IIoT) networks with multiservice and differentiated resource requirements to meet increasing demands through efficient use and management of network resources. Typically, the network slice orchestrator relies on demand forecasts for each slice to make informed decisions and maximize resource utilization. The new generation of Industry 4.0 has introduced digital twins to map physical systems to digital models for accurate decision-making.
In our approach, we first use graph-attention networks to build a digital twin environment for network slices, enabling real-time traffic analysis, monitoring, and demand forecasting. Based on these predictions, we formulate the resource allocation problem as a federated multi-agent reinforcement learning problem and employ a deep deterministic policy gradient to determine the resource allocation policy while preserving the privacy of the slices. Our results demonstrate that the proposed approaches can improve the accuracy of demand prediction for network slices and reduce the communication overhead of dynamic network slicing.
\end{abstract}


\section{Introduction}
The explosive growth of the Internet of Things (IoT) poses significant hurdles for network operators. As a result of the rapid development of communication and sensor technologies paving the way to realize the  IoT, it has become more challenging to support urgent and reliable communications with their quality of service (QoS) requirements \cite{8580366}. As a core component of the 5G architecture, Network Slicing supports various application scenarios and customized services. The physical infrastructure can be divided into logical network instances, called slices\cite{6117098}. However, IoT networks are characterized by considerable differences between service types \cite{9385375}. Hence, the dynamic module for resource allocation should meet individual user requirements. Dynamic network slicing algorithms developed in recent years take advantage of the large amounts of data flowing through the network, which contain information relevant to resource allocation decisions. Using the data generated by the network, the network slices can predict and exploit the upcoming behavior of a system with many different actors. This will enable slices to allocate resources accurately to their users. To this end, accurately predicting the demand for a given network slice is critical to increasing the utilization of network slices.

Some works study the impact of demand forecasting for network slicing and dynamic resource allocation for IoT 5G services, while others analyze slicing as a radio access network (RAN) sharing issue. In \cite{8736403}, a framework is proposed to implement a capacity prediction algorithm that considers guaranteed and best-effort traffic. The authors in \cite{8761246} proposed an AutoRegressive Integrated Moving Average (ARIMA)-based traffic analysis to prevent slice-level agreement violation. In addition, various Deep Q-Network (DQN) algorithms are used to address the problem of demand-based resource allocation, where an agent learns how to perform optimal actions in an environment by observing state transitions and providing feedback \cite{9322106}. As a result of the increasing number of IoT devices and slices within the network, the network cannot handle the workload associated with large networks resulting in scalability issues. Furthermore, for dynamic network slicing to be successful, resource-sharing strategies must be collaborative and autonomous. There is also a need for network slices to share information with the infrastructure providers (InP). This includes user numbers, resource requirements, and any other sensitive information. However, network slices may not want to share because of privacy concerns. Data such as these are stored in a centralized database. Accessing them from the centralized InP would be computationally prohibitive due to the amount of processing and optimization required, as well as privacy issues associated with a centralized data store. Finally, many works have examined demand forecasting for network slices but primarily focused on modeling a single sequence. This is constrained to only account for the time series dependence on traffic networks and ignores the underlying spatiotemporal topological traffic information of the IoT devices in the slice. Therefore, the forecasting model for network slicing should be able to capture the spatial-temporal evolution of traffic generated by the slices.

This paper proposes an intelligent digital twin (DT) framework as a scalable solution for real-time data-driven monitoring and accurate prediction of IoT demand in 5G-based IoT network slices to address the above issues. As an emerging digitization method, DT provides a viable alternative to capture the dynamic and complex network environment \cite{8972134}. It creates a virtual environment in digital space through software definition. By utilizing a graph learning model algorithm, we can construct DTs, i.e., virtual representations of IoT devices with similar structural properties to the physical network. We propose a Graph-Attention-Network (GAT) based DT that can capture the temporal characteristics and accurately monitor and predict the traffic demand of each slice. Based on the demand predictions, we formulate the resource allocation problem as a decentralized multi-agent reinforcement learning (RL) problem to find the optimal resource allocation policy under slice service demand uncertainty. Furthermore, we rely on federated learning (FL), a decentralized machine learning technique, to train deep learning models without compromising privacy. The outcome is a digital twin multiagent federated learning (DT-MAFL) network slicing. The remainder of this article is organized as follows: In Section II, we describe the system model and propose the dynamic resource allocation problem. Section III describes the construction of the DT for each slice. Section IV presents the design of a federated multiagent learning system for dynamic slicing. Section V provides a detailed simulation to validate our proposed slicing framework. Section VI concludes the paper. 
	\begingroup  
	\begin{table}[!b]
		\caption{Summary Of Main Notations}
		\centering
		\smallskip\noindent
		\resizebox{0.8\columnwidth}{!}{%
		 \begin{tabular}{l|l} \hline 
			{Notation }& { Definition} \\  \hline
				$t \in T$ & Time slots\\  \hline 
				$m \in \mathcal{M}$& Number of slices   \\ 	\hline $u \in \mathcal{U}$  &  Number of users \\  \hline   
			    $r_{u, t}^{m}$  & Data rate by user\\  \hline  
		       $\lambda_{u} $ &  Packet arrival rate \\  \hline  
		       $d_{m}$ & Demand of slice $m$ \\  \hline
		       $p_{u, t}$ & Transmit power on base station \\  \hline
		        $h_{u, t}$ & Channel coefficient\\  \hline
		        $\Theta_{u}$ & Shadowing effect  \\  \hline
		        $g_{u}$ &  Rayleigh fast fading \\  \hline
		        $\mathbf{w}_{m}$ &   Amount of RBs for slice $m$  \\  \hline
		        $\varphi_{m}$  &  Number of $\mathrm{RBs}$ needed by slice $m$ \\  \hline 
		        $\tau_{u}$ & The average delay experienced by user $u$\\  \hline 
				$U_u (\cdot)$ & QoS satisfaction  \\  \hline
				$\Omega_{m}^{t}$  &  Resource utilization for a slice \\  \hline 
				$\mathcal{D}$ & Data samples for slice \\  \hline 
				$\mathcal{G}$ & Graph \\  \hline
				$\boldsymbol{q}$ & Weight vector \\  \hline
				$W^{z} $  & Weight   matrix \\  \hline
				$\boldsymbol{A}$ & Adjacency matrix \\  \hline
				 $\beta$ & hyper-parameter \\  \hline
				 $\mathbf{M}^{1}, \mathbf{M}^{2}$ & Dynamic Filters \\  \hline 
			     $\theta$   &  Learning parameter for slice \\  \hline 
				$s$, $a$, $\mathcal{R}$  & State, action and reward of a slice 
			  \\ \hline   
			$\pi$, $y$, $\nabla_{\theta}J$ & Policy , Target and         gradient  policy and for slice $m$   \\ \hline 
				$\eta$  &  Learning rate \\ \hline 
				$\gamma$  &  Discount factor  \\ \hline 
		\end{tabular}}
	\end{table}
	\endgroup 
\section{System Model} 
This section describes the RAN slicing system model for 5G enabled IoT networks. As shown in Fig. 1, the wireless spectrum resources provided by InP are abstracted to constitute a shared resource pool that consists of a certain number of available resource blocks (RBs). These resources can be leased and shared between different network slices. To meet the expected service requirements, the a controller in the cloud is assumed to manage the shared resource pool and allocates resources to slices according to their characteristics \cite{8580366}.
\subsection{Network Model}
We consider an orthogonal frequency division multiplexing (OFDM) system in which a single BS $b$ is assumed to be shared by $\mathcal{M}=\{1,2,3,4 \ldots M\}$ network slices in a downlink setup. Each slice has a set of IoT devices, denoted by $\mathcal{U}$, and can be represented by $\mathcal{U}=\{1,2,3,4 \ldots \mathcal{U}_{\mathcal{M}}\}$.
\begin{figure*}[!t]
		\centering
		\includegraphics[width=\textwidth]{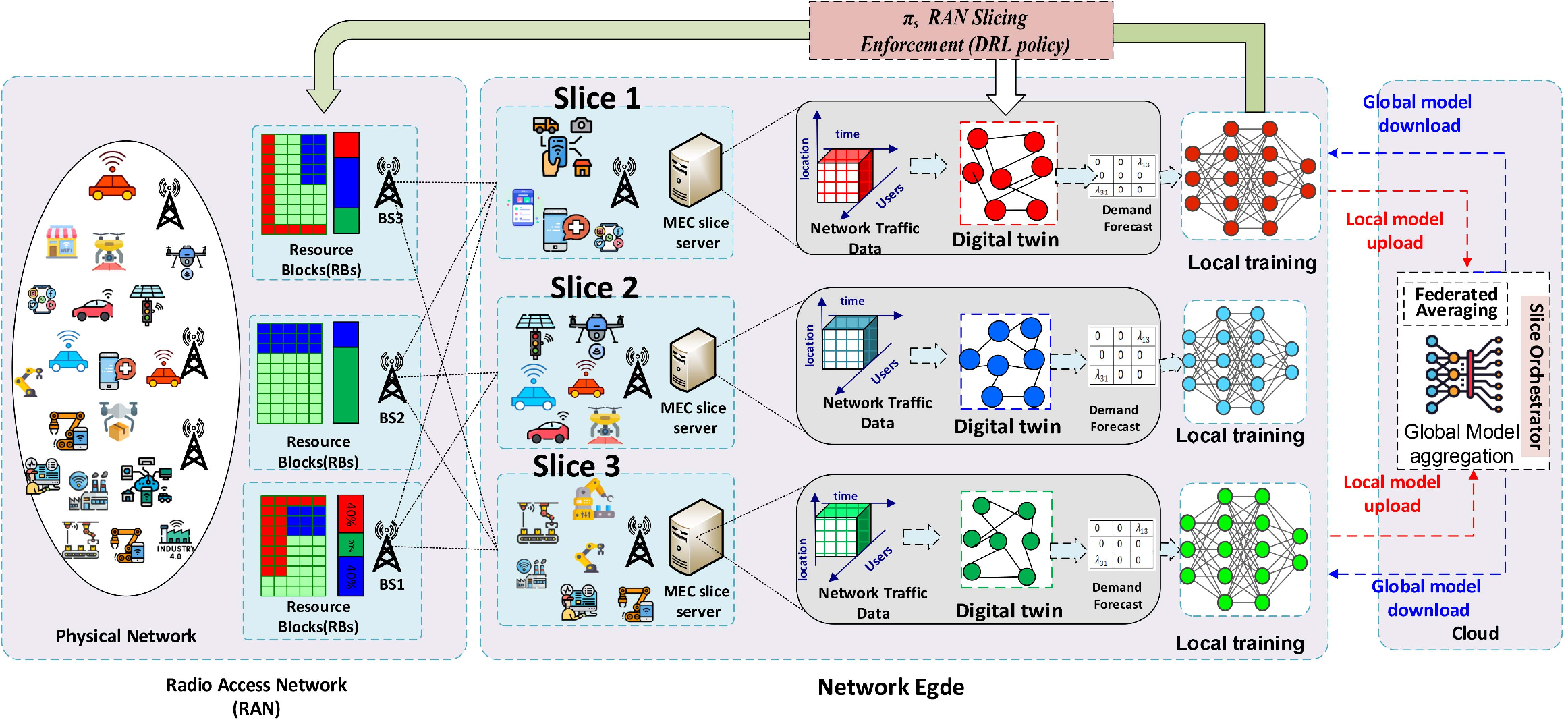} 
		\caption{DT-MAFL dynamic network slicing architecture.} 
		\label{Framework11} 
\end{figure*}
In our network model, time is divided into slots denoted by $t$. Each slice is characterized by a demand ${d}_{m}=\left\{r_{u}, r_{u}, \ldots, r_{u}\right\} \quad(u \in$ $\{1,2, \ldots \mathcal{U}\})$, which is the determinant factor for allocating resources  $w_m$ to slices. In this paper, Shannon theory is used to define the transmission rate for IoT devies, i.e., 
\begin{equation}  
\begin{aligned}
r_{u}=w_{u, t}^{m} \log _{2}\left(1+ \frac{p_{u, t}\left|h_{u, t}\right|^{2}}{N_{0}}\right)
	\end{aligned}
\end{equation}
where $w_{u, t}^{m}$ is the amount of resources assigned to user $u$ and $p_{u, t}$ is the transmit power on the base station at the $t$-th TTI. $h_{u, t}$ is the channel coefficient, which contains path loss, shadowing effect $\Theta_{u}$ and Rayleigh fast fading $g_{u,}$ between IoT device $u$ and the base station. We define as $h_{u, t}= 10^{-\text{PL}^{*}\left(\mathbb{d}_{u,}\right) / 20} \sqrt{\vartheta_{u,} \Theta_{u,} g_{u,}}$ where (PL) is the Path Loss and is defined as $(\mathrm{dB})=20 \log _{10}(d)+20 \log_{10}(f)-$ $27.55$ where $f$ (in $M H z$ ) and $d$ (in meters) representing IoT device to base station channel frequency and distance respectively \cite{9070169}. Moreover, $N_{0}$ is the power of additive white Gaussian noise (AWGN).  The average delay $\tau_{u}$ experienced by the IoT device $u$ in the slice $m$ is $\tau_{u}=\frac{1}{r_{b} - \lambda _{u}}$ where $\lambda _{u} $ is the packet arrival rate. Based on IoT device requirements, different slices can be created to serve different types of IoT device traffic.  A summary of commonly used notations is provided in Table I. 
\subsection{Utility Model}
We introduce a utility function to quantify the QoS needs of IoT devices in a slice \cite{9070169}. We define QoS utility as the satisfaction of user $u$ on either data rate or delay. The utility function of a slice represents the user preference in the network by mapping the achievable rate and maximum delay with the level of UE satisfaction. Here, we use a sigmoid function to express users' satisfaction in terms of rate and delay. In this article, we adopt a framework for satisfaction function $U(\cdot)$ that can meet different optimization goals for both delay and rate constraint as $U\left(\tau_{u}\right)$ and $U\left(r_{u}\right)$, respectively. Therefore, we propose a unified utility function for both delay and rate constraint users.
\subsubsection{Rate-constrained slice}
For the slice $m$, the average QoS satisfaction of a rate-constrained  user is defined as; 
	\begin{equation}  
		U_u \left(r_{u}\right)  = \frac{1}{1 +  e ^{-\phi(r_{u} - r_m^{min})}},
	\end{equation}
where $r_{m}^{min}$ denotes the minimum rate requirement of the user in slice $m$, $\phi$ specifies the steepness of the satisfactory curve.
\subsubsection{Delay-constrained slice} {The average satisfaction on delay in slice $m$ is defined as}
\begin{equation}  
		U_u \left(\tau_{u}\right) = \frac{1}{1 + e^{-\phi(\tau_u^{max} - \tau_u)}}, 
\end{equation}
where the maximum tolerance delay is indicated as $\tau_u^{max}$, which is required to satisfy the upper bound delay for the IoT device $u$. Essentially, a slice utility $	U_m$ can be defined as a summation of individual utilities of the users who make up that slice. The slice utility is defined as:  
	\begin{equation}  
		U_m = \sum_{u \in \mathcal{U}_{n}} U_{u}\left(. \right)
	\end{equation}
Furthermore, the average utilization of a slice is defined as
	\begin{equation}  
		\Omega_{m}= \frac{\mathbf{w}_m}{\varphi_{m}}
	\end{equation} 
where $\mathbf{w}_{m} = \sum_{u \in \mathcal{U}_{m}}w_{u, t}^{m}$ is the amount of RBs occupied by slice $m$ and $\varphi_{m}$  is the number of $\mathrm{RBs}$ needed by slice $m$ which is the demand of the slice. 
\subsection{Federated Learing Model}
Each slice $m$ has a set of data samples and trains its local model using stochastic gradient descent (SGD) \cite{8737464}.  We formulate a learning problem for dynamic slicing  where each slice has the task of optimizing the global loss function $F(\theta)$ by minimizing the weighted average of the local loss function $F_{m}(\theta)$  as 
\begin{equation}   
\min  \left\{F(\theta) \triangleq \sum_{m \in \mathcal{M}} \frac{\left|\mathcal{D}_{m}\right|}{|\mathcal{D}|} F_{m}(\theta)\right\}
\end{equation}
Taking our proposed slicing model into account, we propose that the centralized cloud, i.e. the InP, only performs the aggregation of the model and network slices autonomously perform dynamic network slicing.
\subsection{Problem Formulation}
In our proposed dynamic slicing framework, our goal is to optimize the resource utilization of a slice and ensure the the QoS requirement of the IoT devices in the slice are met. We formulate the optimization problem as maxizing the QoS utility of the slice as:
\begin{equation}   
\begin{aligned}
&\underset{b}{\arg \max }\left\{ \Omega_{m} + U_{m}\right\} 
\\ & \text {s.t. }  \mathbf{w}_{m} \leq \kappa, \quad \quad \mathbf{w}_{m}>0 \\ 
\end{aligned}
\end{equation}
where $\kappa$ is the maximum amount that can be allocated to a slice. $\mathbf{w}_{m} \leq \kappa$ indicates that the amount of resources allocated to the slice should not exceed the total resources available at the base station, and $\mathbf{w}_{m}>0$ are the constraints that ensure that at least one slice always has resources. In dynamic network slicing, not only the current service demand needs to be considered, but the future demands of the slice also need to be considered. Hence, we proposed a DT model which monitors the real-time behavior and accurately predicts each slice's demand.

\section{Adaptive DT based demand  Forecast for RAN slices}
In this section, we design a dynamic graph-based DT to capture the dynamics and spatial dependency of traffic in a slice. During the resource allocation stage of a slice, the graph-based DT is composed of four steps: \textbf{ \textit{i:} Feature Extraction}: a CNN-based feature extraction net is used to convert the raw spatial-temporal data into the feature matrix representations. \textbf{\textit{ii:} Graph Construction:} We adopt a direct optimization approach to learn dynamic graph structures, which generate a graph adjacency matrix that represents the DT.  \textbf{\textit{iii:} Interaction Modeling:} Captures the dynamic interaction of UEs and the structural information within the nodes (UEs and gNodeB) of the slice. \textbf{\textit{iv:} Prediction Model:} integrates each node's individual sequential information to predict the slice's traffic demands. The network topology of the slice at each time interval can be represented as a sequence of graph snapshots $\mathcal{G}=\left\{\mathcal{G}_{1}, \cdots, \mathcal{G}_{T}\right\}$, where $T$ is the number of time steps.

To construct the DT, the network topology of the IoT devices in the slice is represented as a graph. The graph can be defined as $\mathcal{G}=\{\mathcal{V}, \mathcal{E}\}$ with $N$ nodes, where $\mathcal{V}$ and $\mathcal{E}$ are the set of nodes and edges, respectively. The network traffic data generated by the slice is defined as $d \in \mathbb{R}^{Z \times T \times V}$ where $Z$ is the input channels and $T$ time intervals and $V$ are the spatial data associated with the channel \cite{8999421}. The task of traffic forecasting is to learn a mapping function $h(\cdot)$, which takes historical traffic data $d$ and graph $\mathcal{G}$ as inputs to forecast future time intervals traffic data:
\begin{equation}   
\hat{d}_m=h(d_m , \mathcal{G}, \psi),
\end{equation} 
where $\psi$ is the learnable parameters. However, the Slice traffic data is defined as multivariate and temporal data and no nodes. Therefore, the proposed DT is used to build a graph that comprehensively models the interaction between IoT devices in a slice.
\subsection{Graph construction.} 
The first stage of the proposed slice DT is a graph learning layer that learns a graph adjacency matrix adaptively to capture the hidden relationships among time series data for the slice. The adjacency matrix can be seen as DT, representing the topology and spatial behavior of the IoT devices in a slice. Firstly, CNN-based feature extraction is used to convert the raw spatial-temporal data into the feature matrix representation $\mathbf{B}^{m}$. Based on the feature matrix,  a unique graph structure is generated for the slice, consistent with the dynamic property of the traffic data generated by BS. Inspired by \cite{3403118}, the underlying adjacency matrix $\mathbf{A}$ is dynamically generated with:
\begin{equation}   
\begin{aligned}
\boldsymbol{A} =\operatorname{ReLU}\left(\tanh \left(\beta\left(\mathbf{M}_{m}^{1} \mathbf{M}_{m}^{2}{ }^{\mathrm{T}}-\mathbf{M}_{m}^{2} \mathbf{M}_{m}^{1}\right)\right)\right)
\end{aligned}
\end{equation} 
where $\beta$ is a hyper-parameter which is a saturation rate of the activation function and $\mathbf{M}^{1} = \mathbf{M}^{2}=\tanh \left(\beta (\mathbf{E}_{m}\mathbf{B}^{m}\right)$ is a dynamic filter with $\mathbf{B}^{m}$ as its input.
\subsection{Interaction Modeling for slices with GAT}
The GAT layer can model the relationships between nodes and frequencies \cite{velivckovic2017graph}. The GAT generates the attention weight that reflects the channel quality of traffic generated by IoT devices in a slice. The input to the graph layer is a set of node features $\left\{\boldsymbol{x}_{1}\ldots, \boldsymbol{x}_{N}\right\}$ embedded in the adjacency matrix $\boldsymbol{A}$ and the output is a new set of node features $\left\{\boldsymbol{x}_{1}^{\prime}, \ldots, \boldsymbol{x}_{N}^{\prime}\right\}, \boldsymbol{x}_{z}^{\prime} \in \mathbb{R}^{H}$, where $H$ is the number of features for each node. We compute the attention coefficient of node $z \in \mathcal{N}_{z}$ to node $v$ as follows \cite{velivckovic2017graph}:
\begin{equation}  
\begin{aligned}
\alpha_{z v}=\frac{\exp \left(\text{LeakyReLU}\left( \boldsymbol{q}^{T}\left[W^{z} \boldsymbol{x}_{z} \| W^{z} \boldsymbol{x}_{v}\right]\right)\right)}{\sum_{w \in N_{v}} \exp \left(\text{LeakyReLU}\left(\boldsymbol{q}^{T}\left[W^{z} \boldsymbol{x}_{u} \| W^{z} \boldsymbol{x}_{v}\right]\right)\right)}    
\end{aligned} 
\end{equation}
where $\mathcal{N}_{v}$ is the set of immediate neighbors and $W^{z} \in \mathbb{R}^{F \times D}$ is the weight matrix of the graph. $\boldsymbol{q} \in \mathbb{R}^{2 D}$ is a weight vector of each node on the graph. Note that $A_{zv}$ is the weight of the link $(u, v)$ in the current snapshot $\mathcal{G}$. $\alpha_{zv}$ indicates the importance of node features in traffic data. The attention coefficients are then used to compute the final output features for every node:
\begin{equation}  
\begin{aligned}
\boldsymbol{x}_{z}^{\prime}=\sigma\left(\sum_{u \in \mathcal{N}_{v}} \alpha_{z v} {W}^{s} \boldsymbol{x}_{v}\right)
\end{aligned} 
\end{equation}
where $\sigma(\cdot)$ is applied component-wise. The GAT layer is able to dynamically learn the relationships between different channels and timestamps and abstract more meaningful information from the traffic channel of the slice. 

\subsection{Prediction Model}
Finally, a two-layer fully connected neural network is used to obtain the final traffic prediction as follows:  
\begin{equation}  
\begin{aligned}
\hat{d}_{m}=\operatorname{softmax}\left(\boldsymbol{x}_z^{\prime}  \psi+\boldsymbol{b}\right)
\end{aligned} 
\end{equation}
where $\psi$ is a learnable matrix and $\boldsymbol{b}$ is the bais. We define a cross-entropy loss to train the model as:
\begin{equation}  
\operatorname{Loss}_{\text {forecasting }}(t)=\frac{1}{T} \sum_{t=1}^{T}\left|d_{m}^{(t)}-\hat{d}_{m}^{(t)}\right|
\end{equation} 
where ${d}_{m}^{(t)}$ and $\hat{{d}}_{m}^{(t)}$ are the actual demand and the predicted demand of the slice $m$ at timestamp $T$.

\section{Multi-Agent Federated Learning Dynamic Network Slicing}
The limited communication capability of wireless networks makes DQN solutions unsuitable for large networks. We propose a multi-agent federated dynamic slicing algorithm to ensure privacy and scalability. The proposed algorithm can determine the near-optimal slicing policy that meets the QoS satisfaction requirements. According to the result of the prediction of the DT forecast, the dynamic slicing algorithm aims to maximize the average resource utilization of each slice and user satisfaction. Specifically, we model FL with multiple slices as an MDP and design a multi-agent federated reinforcement learning algorithm to explore the optimization solution for IoT slices. 
\subsection{Problem Transformation}
To maximize the long-term return for the provider while accounting for the real-time arrivals of IoT network slices, we transform the problem into a Markov decision process (MDP) \cite{9322106}. An MDP is defined by a tuple $\left\langle t, \mathcal{S}, \mathcal{A}, R\right\rangle$ where $t$ is an decision epoch, $\mathcal{S}$ is the system's state space, $\mathcal{A}$ is the action space and $R$ is the reward function. The state, action, and reward can be defined as follows.
\paragraph{State} At the beginning of each time interval, the state of each slice is defined as $ {s}_m(t)=\left[d_m(t), \hat{d}_m(t-1)\right],$ where ${d}_{m}^{(t)}$ and $\hat{{d}}_{m}^{(t)}$ are the actual demand and the predicted demand of the slice $m$.
\paragraph{Actions} The action taken by the slice $m$ at time $t$ is represented by $a_{t}^{m}=\left\{w_{m}\right\}$, where $w_{m}$  represents the increase in the percentage of bandwidth for the $m$-th slice. When $w_{m}>0, w_{m}$ percentage of bandwidth will be added to the rsource of the slice and when $a_{m}<0,\left|a_{m}\right|$ percentage bandwidth of the slice will be reduced.
\paragraph{Reward}
After the state transition, each slice would gain rewards w.r.t. current state $s_{m}^{t}$ and actions $a_{m}^{t}$. The reward function of the of the slice is defined as:
\begin{equation} 
	\begin{aligned} \label{reward}
\mathcal{R}_{m}^{t}(s, a)=   \left\{\Lambda \cdot \Omega_{m}^{t}(s,a) + \mu \cdot U_{m}(s,a)\right\}
	\end{aligned}
\end{equation} where $\Lambda$ and $\mu$ are the importance of the algorithm  places on the resource utilization and utility, respectively \cite{9070169}. The ultimate objective of the agent in the multi-agant FL system is to find the optimal slicing strategy (policy) $\pi^{*}$.
\subsection{MAFL Network slicing} 
Network slices collaborate to find the optimal resource allocation in a decentralized manner. The role of the slice orchestrator is to coordinate all slices to achieve a global optimal resource allocation. In this allocation process, network slices do not need to share or exchange their proprietary information, including resource availability and traffic dynamics, with each other or with the slice orchestrator. the MAFL Network slicing, consist of 1) Local Policy Iteration 2) Global Policy Aggregation. The local policy iteration is performed by the slices and the central InP controller performs the  for global policy aggregation. The aggregation period for the global model is called $\tau$.
\begin{center}
	\begin{algorithm}[!b]  
		\caption{{  Multi-Agent Federated Learning  Dynamic Network Slicing}}	
		Initialize all slices  with the same model, i.e., $\theta^{(m)}=$ $\theta$\;
		\For{ $t=1,2, \ldots, T$}{
			\textit{$\triangleright$ Local Policy Iteration}\;
			\For {each slice $m \in \mathcal{M}$ in parallel}{ 
				Obtain the current observation $d , \hat{d}$  from the DT\; 
				Select a random action $a^{t}_{m}$ with probability $\epsilon$\;
				Otherwise, select action $a^{t}_{m}$ satisfying $a_{t}=_{a_{t} \in A}{\operatorname{argmax}} Q\left(s_{t}, a_{t}\right)$\;
				Update the resource fraction $\mathbf{w}_{m}$ and observe reward $R_{m}$ and new state $s^{t+1}$ \;
				Store experience $\left\langle s^{t}_{m}, a^{t}_{m}, \mathcal{R}_{m}, s^{t+1}\right\rangle$ and sample mini-batch  $\mathcal{D}$\;
				Update the local model as $\theta_{t}^{(m)}$ according to (\ref{local})\;
				Update $\theta_{t}^{(m)}$ as $ {\theta}$.\;
				Update the main actor network using the sampled policy gradient according to (\ref{gd}) and target network based on  (\ref{eq:20})
			}
			$\triangleright$ \textit{Global Policy Aggregation}\;
			\If{$\bmod \left(t, \tau \right)=0$}{
				Collect the most updated model from the slices\;
				Obtain ${\theta}$ by performing model aggregation according to (\ref{global})\;
				Broadcast ${\theta}$ to to all slices\;
			}
		}
	\end{algorithm}
\end{center} 
\begin{figure*}[!hb]
		\centering
		\begin{minipage}{\textwidth}
			\centering
			\begin{subfigure}[b]{0.45\textwidth}
				\centering
				\includegraphics[width=\textwidth]{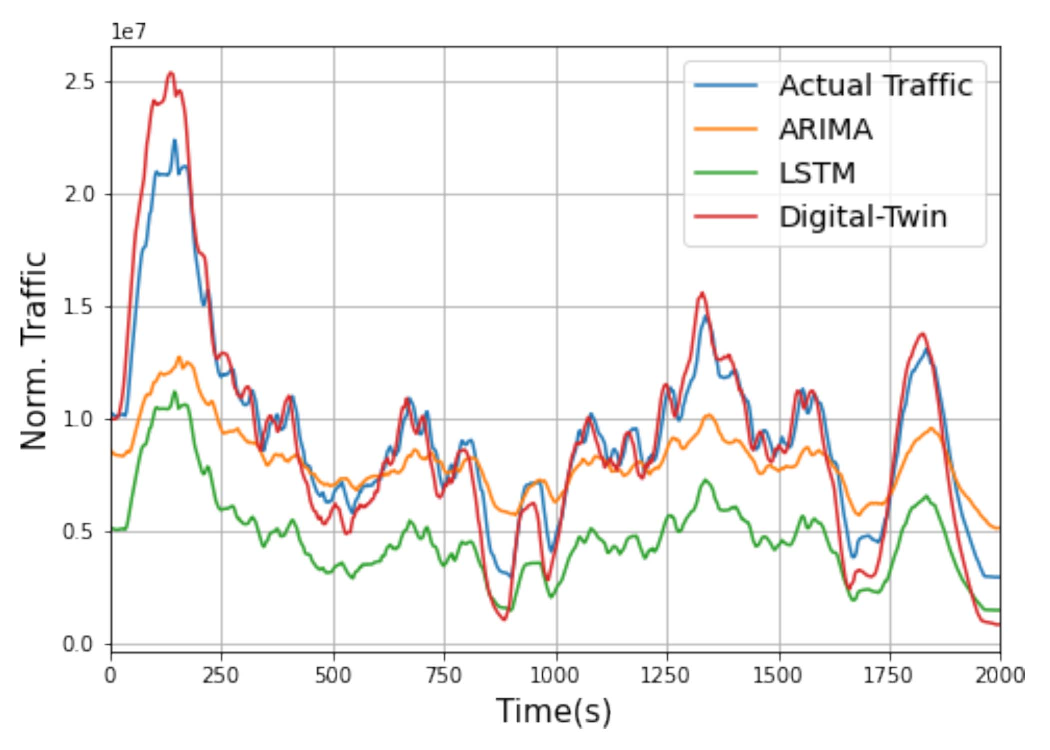}
				\caption{Demand forecast}\label{fig:2a} 
			\end{subfigure}%
			\begin{subfigure}[b]{0.45\textwidth}
				\centering
				\includegraphics[width=\textwidth]{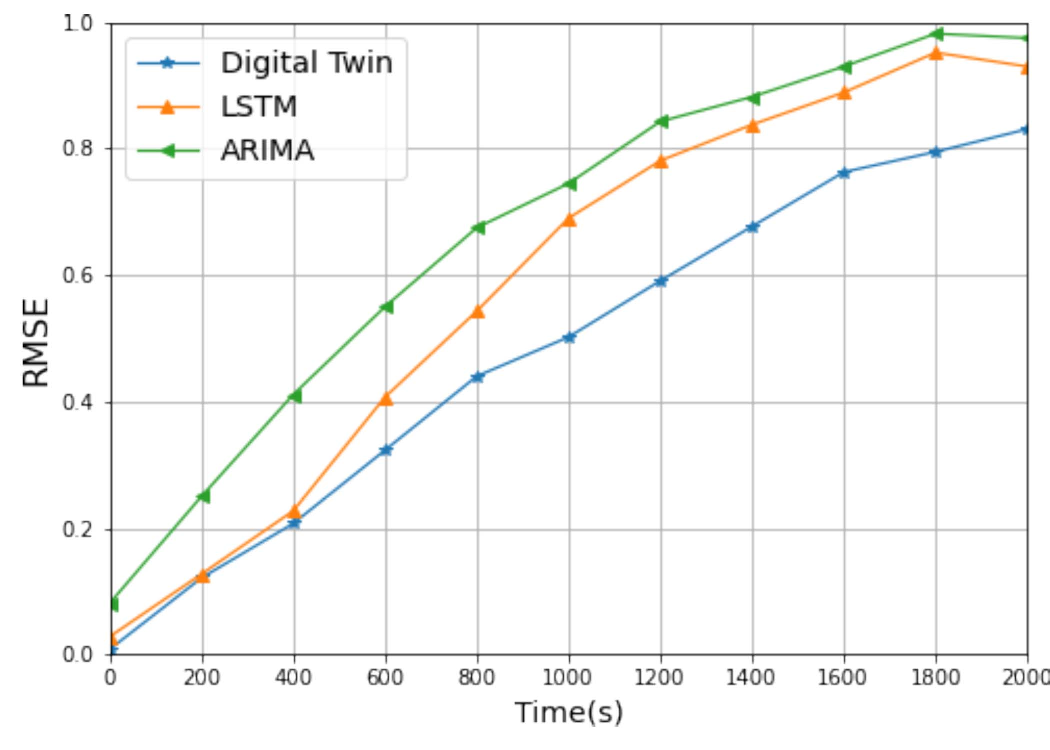}
				\caption{Prediction errors on RMSE}\label{fig:2b}
			\end{subfigure}%
			\caption{\textit{Forecasting Performance by various models.}}
			\label{fig:2}
		\end{minipage}
	\end{figure*} 
\subsubsection{Local Policy Iteration} Taking advantage of Deep Deterministic Policy Gradient (DDPG) \cite{9322106}, an actor's network guides the update of policy parameters $\theta_{t}$ of the network slcing policy $\pi_{\theta_{l}}$ based on the evaluated values from the critic's network. The policy gradient is defined as follows:
\begin{equation}  
	\begin{aligned} \label{gd}
\nabla_{\theta \pi} \approx \frac{1}{n} \sum_{m} \nabla_{a} Q\left(s_{m}, a \mid \theta^{Q}\right) \nabla_{\theta^{\pi}} \pi\left(s_{m} \mid \theta^{\pi}\right) .
	\end{aligned}
\end{equation}
We update the critic by minimizing the loss:
\begin{equation}  
	\begin{aligned}
\mathcal{L}\left(\theta^{Q}\right)=\frac{1}{n} \sum_{m}\left(y_{m}-Q\left(s_{m}, a_{m} \mid \theta_{Q}\right)\right)^{2}
	\end{aligned}
\end{equation}
where $ y_{m}=\mathcal{R}_{m}+\gamma Q^{\prime}\left(s_{i+1}, \pi^{\prime}\left(s_{i+1} \mid \theta^{\pi^{\prime}}\right)\right)$. The target parameters in both actor and critic networks are updated as follows:
\begin{equation} \small
\label{eq:20}
\begin{aligned}
&\theta_{i}^{\pi^{\prime}} \leftarrow \nu \theta_{i}^{\pi}+(1-\nu) \theta_{i}^{\pi^{\prime}} \\
&\theta_{i}^{Q^{\prime}} \leftarrow \nu \theta_{i}^{Q}+(1-\nu) \theta_{i}^{Q^{\prime}}
\end{aligned}
\end{equation}  

In the $t$-th iteration, each slice agent performs model update via the stochastic gradient descent (SGD) algorithm based on its local data with the following expression:
\begin{equation}  
	\begin{aligned} \label{local}
\theta_{t}^{(m)} \leftarrow \theta_{t-1}^{(m)}-\eta \nabla f_{m}\left(\theta_{t-1}^{(m)}\right)
 \end{aligned}
\end{equation}
where $\eta$ is the learning rate and where $f_{m}(\theta)=$ $\frac{1}{\left|D_{m}\right|} \sum \mathcal{L}_{m}\left(\theta^{m}\right)$ is the loss function of the agent $m$, $\theta_{t-1}^{(m)}$ is the local model of the slice at the start of the $t$-th iteration. The local model update $\theta_{t}^{(m)}$ is uploaded to the slice orchestrator for global model aggregation.
\subsubsection{Global Policy Aggregation} The slice orchestrator updates the global network slicing model by aggregating all local model updates from IoT slices as follows:
\begin{equation}  
	\begin{aligned} \label{global}
{\theta} \leftarrow \sum_{m \in \mathcal{M}} \frac{\left|\mathcal{D}_{m}\right|}{\left|\mathcal{D}\right|} \theta_{t}^{(m)}
 \end{aligned}
\end{equation}
where ${\theta}$ denotes the aggregated model at the cloud server. After that, ${\theta}$ is broadcasted to the slices, which can be expressed $\theta_{t}^{(m)} \leftarrow {\theta}$. The global policy ${\theta}$ is used in the next training period. Our proposed algorithm, summarized in Algorithm 1, can find the optimal secure resource allocation strategy without sharing information and experience. The communication cost is analyzed as the total number of messages transmitted between the client and server. Taking into account the costs of forward and backward propagation, the entire network requires $O\left(l k^{2}\right)$ communication messages for an iteration where $l$ is the number of layers and $k$ is the number of neurons. 

\section{Numerical Evaluations} 
\subsection{Simulation Environment Settings}
The system consists of a base station with a coverage radius of 500 meters and a carrier frequency of 2 GHz. The average number of IoT devices connected to a slice is 100, which are uniformly distributed. We set the number of slices $|\mathcal{M}|$ to 6. We set the bandwidth at 10 MHz and divide it into 50 RBs. The power spectral density of additive white Gaussian noise (AWGN) and interference thresholds are -174 dBm/Hz and -101.2 dBm, respectively. The Path Loss (PL) is defined as PL $(\mathrm{dB})=20 \log _{10}(\mathbf{d})+20 \log _{10}(f)-$ $27.55$ where $f$ (in $M H z$) and $\mathbf{d}$ (in meters) represent IoT device to base station channel frequency and distance respectively \cite{7499297}.  Our DQN model consists of 6 hidden layers with 64 neurons in each hidden layer. We set the discount factor for both the actor and critic neural networks at 0.95 and the learning rate at 0.1. The replay buffer size is 1000, and the minibatch size for sampling is 64. Furthermore, we adopt Adamoptimizer to optimize the loss function \cite{9005674}. TensorFlow 2.0 is used to implement DQN and FL \cite{tensorflow2015-whitepaper}. We compare our proposed DT with two different prediction models, i.e., a vanilla Long short-term memory (LSTM) model \cite{8884230} and ARIMA model \cite{8761246}. Furthermore, our proposed dynamic slicing, i.e., DT-MAFL slicing, is compared with the following benchmarks: (1) Federated Learning-based slicing (FL), which performs resource allocation and sends updates to the slice orchestrator. (2) Multi-agent DQN (MADQN), which models the resource allocation problem as a distributed problem and deploys DQN agents on each slice to allocate resources independently \cite{9322106} and (3) Netshare, which is a centralized controller with a global network view and performs resource allocation decisions for all slices \cite{6117098}. The summary of the simulation parameters used can be found in Table II.
	\begingroup  
	\begin{table}[!t]
		\caption{Simulation parameters}
		\centering
		\smallskip\noindent
		\resizebox{0.8\columnwidth}{!}{%
			\begin{tabular}{|l|l|} \hline 
				{Parameters }& { Values} \\ 	\hline 
				\hline {Cell radius }& {500m} \\  
				\hline {Bandwidth}& {10 MHz} \\  
				\hline {Number of RBs}& {50} \\ 
				\hline {Bandwidth of per RB}& {180kHz} \\  
				\hline {The number of BSs}& 1 \\
				\hline {The number of slice tenants }&{6}\\ 
				\hline {Transmitting power}& {30dBm }\\ 
				\hline {Channel gain}& {path loss model }\\ 
				\hline {Noise Spectral Density }&{-174 dBm/Hz} \\  
				\hline {User location}& {Uniform distribution} \\
				\hline {Mini-batch size }&{64 }\\  
				\hline {Experience replay buffer size }& {1000} \\  
				\hline {Discount factor}&{0.5} \\ 
				\hline {Learning rate }&{0.1 }\\  \hline  
		\end{tabular}}
	\end{table}
	\endgroup 
\begin{figure*}[!t]
	\centering
	\begin{minipage}{\textwidth}
		\centering
		\begin{subfigure}[b]{0.45\textwidth}
			\centering
			\includegraphics[width=\textwidth]{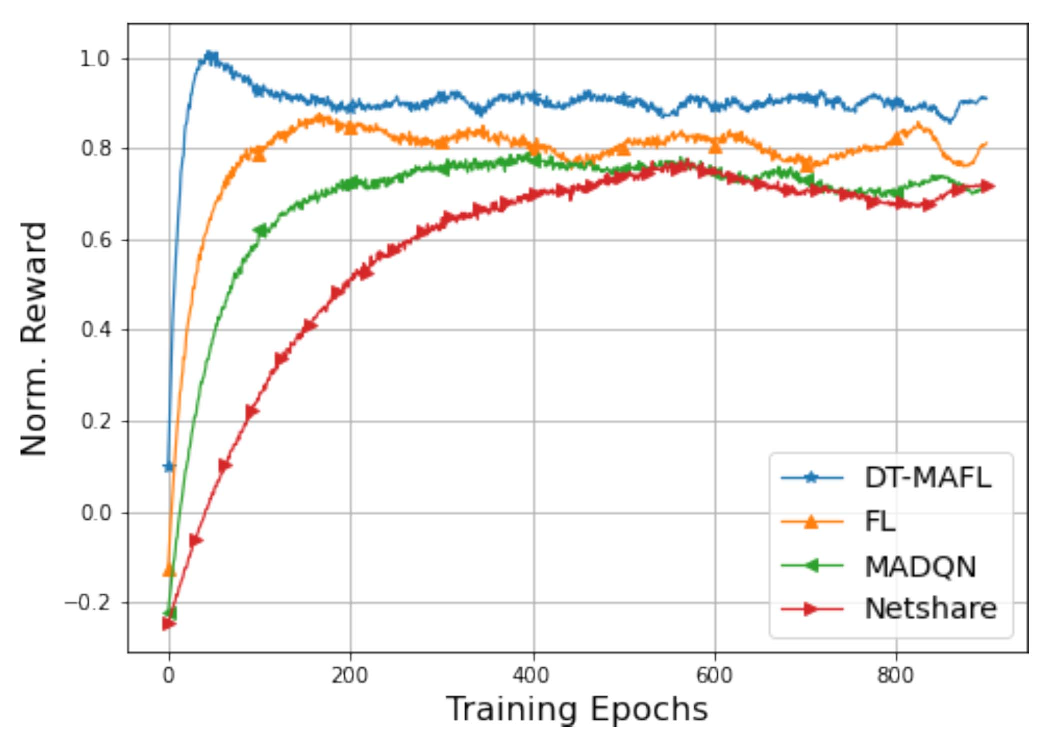}
			\caption{Reward}\label{fig:3a} 
		\end{subfigure}%
		\begin{subfigure}[b]{0.45\textwidth}
			\centering
			\includegraphics[width=\textwidth]{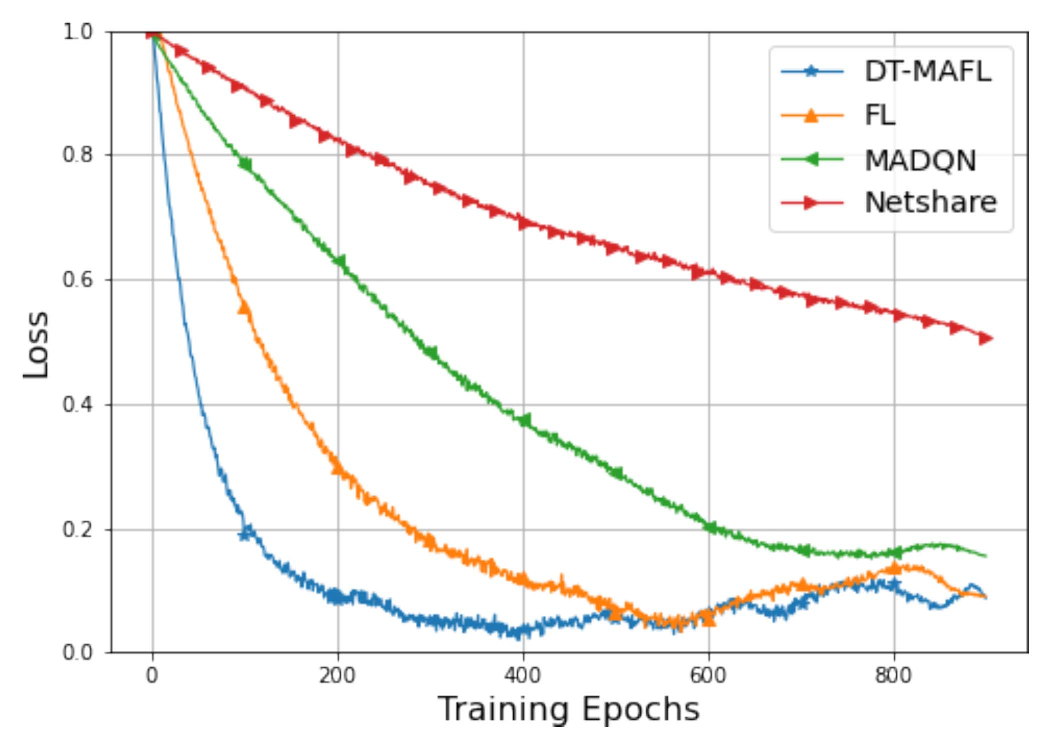}
			\caption{Loss}\label{fig:3b}
		\end{subfigure}%
		\caption{\textit{Convergence performance.}}
		\label{fig:3}
	\end{minipage}
\end{figure*}   
\begin{figure*}[!b]
	\centering
	\begin{minipage}{\textwidth}
		\centering
		\begin{subfigure}[b]{0.45\textwidth}
			\centering
			\includegraphics[width=\textwidth]{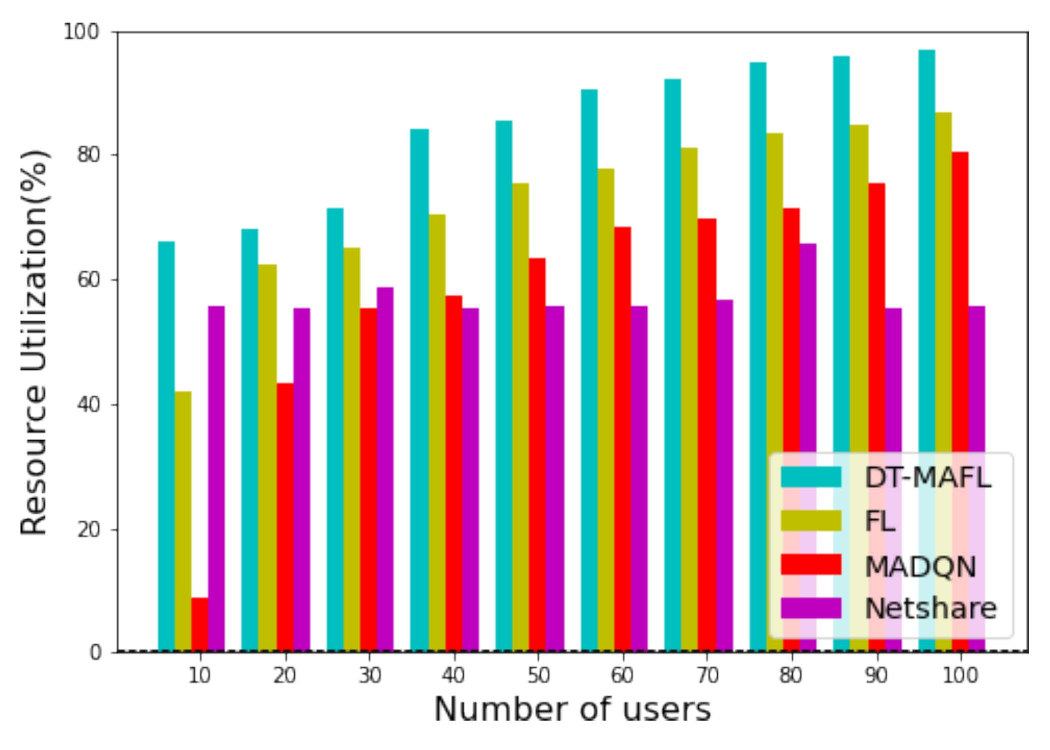}
			\caption{Resource Utilization}\label{fig:4a} 
		\end{subfigure}%
	\begin{subfigure}[b]{0.45\textwidth}
			\centering
			\includegraphics[width=\textwidth]{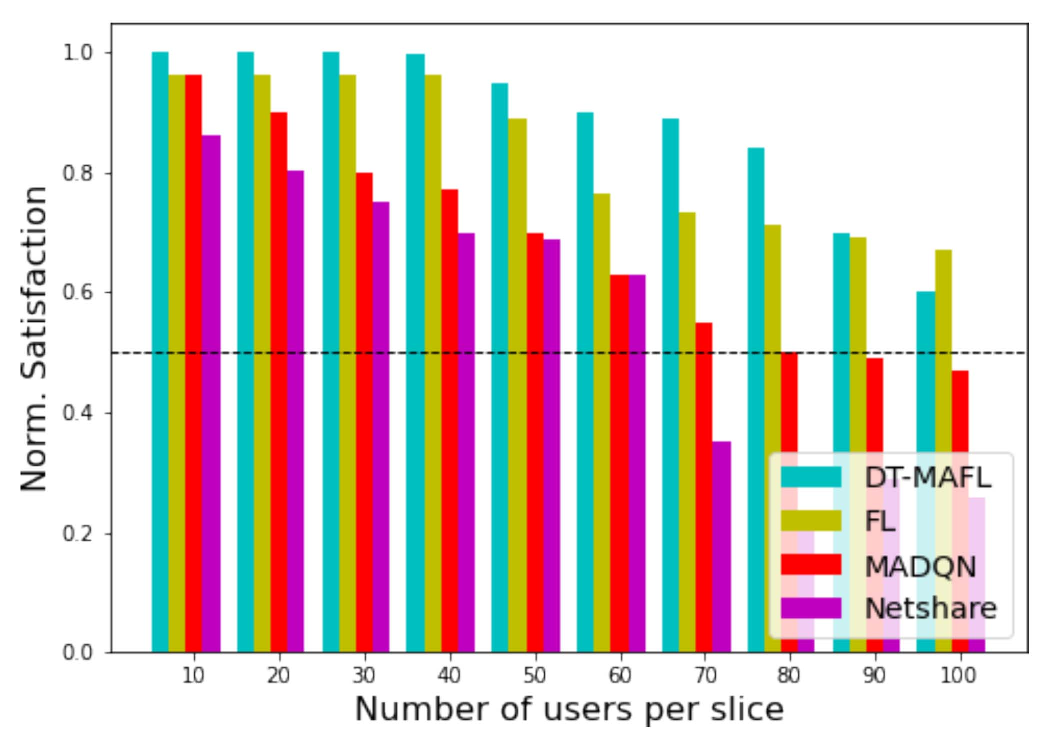}
			\caption{QoS Satisfaction}\label{fig:4b}
		\end{subfigure}%
		\caption{\textit{Resource Utilization and QoS satisfaction performance.}}
		\label{fig:result4}
	\end{minipage}
\end{figure*}
\subsection{Forecasting Performance} 
In Fig. 2, we show the qualitative comparison of the prediction in terms of the average traffic load (Mbs) of each slice in the BS over a period of time. The results show that the DT models can follow the wave of traffic relatively well. The LSTM and ARIMA models cannot follow it, and the ARIMA model has the worst pattern. The GAT network in our proposed DT model encodes spatial features in graph learning, which can track the spatial correlations in the 5G traffic data generated by the IoT devices in a slice. The ARIMA model increases linearly and leads to overfitting of the data set as the number of data increases. In Fig. 2b, the Root Mean Square Error (RMSE) measures the performance of the predictive models. It is helpful to estimate which model carries the largest error point based on RMSE because it magnifies the relative difference between the errors. As a measure of accuracy, the RMSE is a good indicator. Our DT model methods achieve low error rates, highlighting the importance of modeling spatial correlations in traffic forecasting.

\subsection{Convergence Performance}
Fig. \ref{fig:3} illustrates the reward and loss results of our proposed dynamic slicing system. The reward is numerical feedback that evaluates the allocation of resources of the proposed framework in Equation (\ref{reward}). Fig. \ref{fig:3a} shows that our proposed DT-MAFL slicing framework has the highest reward and achieves stability quickly. The FL model can achieve relatively good performance, but the resource allocation is unstable without an accurate prediction model. The MADQN converges slower than FL and DT-MAFL due to the communication overhead. At the same time, the optimal model cannot adapt to the dynamic conditions of the IoT devices in the slice. The loss curve further proves the convergence performance of the proposed scheme in Fig. \ref{fig:3b}. The proposed DT-MAFL has the lowest loss, indicating stable convergence.

\subsection{Performance on Slice Satisfaction and Resource Utilization} 
This section analyzes the system's performance regarding resource utilization and QoS satisfaction. In Fig. \ref{fig:4a}, we compare the resource utilization of the algorithms with the increasing number of users in each slice. From Fig. \ref{fig:4a}, it can be seen that the proposed DT-MAFL achieves a higher allocation of resources than FL, MADQN and Netshare. With DT-MAFL, the prediction model can accurately predict the traffic load for the next time. Therefore, it can adapt to the dynamic changes in slice requests in a 5G network and increase resource utilization. As a result, the QoS requirements of the IoT devices in the slices are satisfied, as shown in Fig. \ref{fig:4b}. Netshare is the most inefficient due to its lack of learning ability. This is because the QoS satisfaction values do not reach the threshold when the number of users in the slice increases. In Fig. \ref{fig:4b}, we can also see that the QoS satisfaction of FL and MADQN decreases rapidly as the number of network users increases. The QoS satisfaction of MADQN in high load scenarios also does not reach the QoS threshold of the slices.

\section{Conclusion}
This paper presents a comprehensive approach to optimize IoT network slice management through advanced traffic analysis and resource allocation. We introduced a novel DT architecture leveraging GAT for real-time traffic monitoring and demand forecasting within network slices. Building upon these predictions, we developed a federated multi-agent reinforcement learning framework for dynamic network slicing, enabling inter-slice collaboration while preserving individual slice privacy. Our extensive simulations demonstrate the efficacy of the proposed methods in significantly enhancing demand prediction accuracy for network slices and substantially reducing the communication overhead associated with dynamic network slicing. These improvements translate to more efficient resource utilization, improved QoS, and enhanced scalability of IoT networks.
Future work includes exploring the integration of policy distillation methods, where knowledge from multiple specialized policies can be consolidated to improve collaboration and address Non-IID (Non-Independent and Identically Distributed) challenges in network slicing due to diverse requirements for slices. This approach could potentially further optimize the learning process and improve the adaptability of our system to diverse network conditions. By leveraging policy distillation, we aim to create more robust and generalizable policies that can effectively handle the heterogeneous nature of network slices, leading to more efficient resource allocation and improved overall network performance.
\section*{Acknowledgment}
This work is supported in part by the Natural Science Foundation of China under Grant No.61806040, Grant No. 61771098, Grant No. 61971079, and Grant No. U21A20447; in part by the fund from the Department of Science and Technology of Sichuan Province under Grant No. 2020YFQ0025; in part by the fund from Intelligent Terminal Key Laboratory of Sichuan Province under Grant No. SCITLAB-1018, and by the Science and Technology Research Program of Chongqing Municipal Education Commission (KJZD-k202000604).
\bibliographystyle{IEEEtran}
\bibliography{main} 
\end{document}